\documentclass[%
superscriptaddress,
groupedaddress,
showpacs,
 amsmath,amssymb,
 aps,
pra,
floatfix,
]{revtex4-2}

\usepackage{graphicx,url}

\begin{document}

\title{Pulse compression by photoexcitation-induced dynamics of Bragg mirrors}

\author{Z. Ruziev}
\affiliation{National University of Uzbekistan,\\ University street 4, 100174\\
Tashkent, Uzbekistan}

\author{K. Yabana}
\affiliation{Tsukuba University, 1 Chome-1-1 Tennodai, \\Tsukuba, Ibaraki 305-8577, Japan}

\author{A. Husakou}
\email{gusakov@mbi-berlin.de}
\affiliation{Max Born Institute, Max Born Str. 2a, D-12489 Berlin, Germany}

\begin{abstract}
We propose dynamical Bragg mirrors as a means to compress intense short optical pulses. We show that strong-field photoexcitation of carriers changes the refractive index of the layers and leads to motion of the resonance-defined boundary of the Bragg mirror. In a reflection geometry, this counter-propagating motion leads to significant compression of the incident pulse. We utilize a finite-difference time-domain numerical model to predict up to a 6-fold pulse compression in the few-femtosecond regime. Modification of the refractive index and properties of the compressed pulse as a function of the incident pulse parameters are investigated.  
\end{abstract}

\maketitle

\section{Introduction}
A plethora of fields of modern ultrafast optics, such as tracing of atomic motion in molecules \cite{zewail}, chemistry on electronic timescale \cite{remacle}, electrooptical sampling \cite{dima}, attosecond streak camera \cite{a-streaking}, photoconductive sampling \cite{nps}, steering of ultrafast electron dynamics in the valence band \cite{remacle2}, material modification \cite{mod} and so on, require short, sub-10-fs intense pulses. The multifarious nature of these phenomena determine the wide range of the required pulse parameters, which in turn requires various methods and techniques for their generation.

In the visible and near infrared ranges, ultrashort pulses are routinely produced using approaches such as directly from a laser oscillator \cite{Ell}, by a non-collinear optical parametric amplifier \cite{Brida}, or by spectral broadening in a nonlinear medium \cite{Nisoli}. Ultraviolet pulses are typically generated in a two-step approach: first, few-optical-cycle pulses in the visible or near-infrared  ranges are generated, and then nonlinear frequency up-conversion is used to reach the UV range \cite{last}. Ultraviolet pulses were also generated by second-order nonlinear frequency conversion in crystals \cite{Varillas}. 

Beyond that, there are other methods of short pulse generation such as non-solitonic radiation during soliton fission \cite{short_hh,Brahms}. Another approach relies upon ultrafast modification of the material properties, such as short pulse emission by transient plasmonic resonance \cite{our_short}.

A constant demand for the novel pulse generation techniques makes the development of new pulse compression methods an important and challenging aim. In particular, the approaches for the generation of intense (above roughly 30 TW/cm$^2$) short (few-fs) near-uv pulses are currently limited. Therefore in this paper we propose a pulse compression method which builds upon recent developments in the ultrafast control of bound-electron and atomic motion \cite{hassan,hoegen} and photoinduced strong modification of dielectric properties within one optical cycle \cite{schultze,hanus,schiffrin}. This approach is also distantly related to the high harmonic generation by oscillating plasma mirror\cite{plasma}. Our proposal is based on photoexcitation-induced dynamics of Bragg mirrors, as explained later in Section 2 and Fig. 1. We simulate strong-field photoexcitation of carriers by a short optical pulse, which changes the refractive index of the Bragg mirror layers. We predict that this leads to motion of the resonance-defined boundary of the Bragg mirror and to a significant compression of the incident pulse.

The paper is organized as follows: In Section 2, we describe the general geometry and the operating principle of the proposed setup, and in Section 3 we provide details of the numerical model. In Section 4, we present and discuss the results of the numerical simulation, followed by a conclusion in Section 5.

\section{Proposed setup and principle of operation}

\begin{figure}
    \centering
    \includegraphics[width=0.5\linewidth]{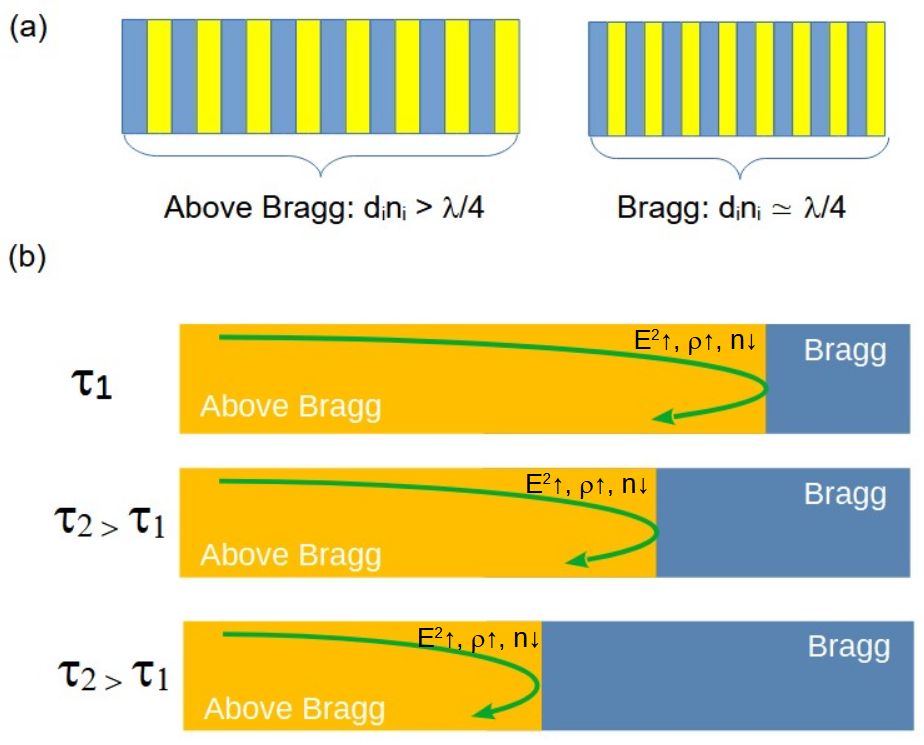}
    \caption{Scheme of the proposed setup. In (a), above-Bragg (left) and Bragg areas (right) are depicted. In (b), the dynamics of the Bragg mirror boundary and of the pulse propagation are illustrated.}
    \label{1}
\end{figure}

Upon reflection from a counter-propagating mirror moving with velocity $v$, an optical pulse would be compressed by a factor $1/(1-v/c)$, where $c$ is the light velocity. Such compression is obviously impossible with the mechanically moving mirror, as it cannot reach velocity comparable to $c$. To circumvent this challenge, we design a system characterized by a time-dependent reflecting boundary, based on Bragg mirrors.

Consider an array of plane-parallel layers of two alternating materials, as depicted in Fig. 1(a), characterized by thicknesses $d_i$ and refractive indices $n_i$, where $i$ is the layer number. If the Bragg condition $n_i=\lambda/(4d_i)$ is met (or nearly met), such as in Fig. 1(a) right, the light with wavelength $\lambda$ will be reflected from the corresponding Bragg layer due to constructive interference of the reflected waves; for brevity, we will call area occupied by such layers as "Bragg" area. On the other hand, if the refractive index $n_i$ is significantly above the $\lambda/(4d_i)$, such as in Fig. 1(a) left, then no strong reflection will be observed; for brevity, we will call area occupied by such layers as "above-Bragg" area.

Now consider a structure which consists of an above-Bragg area followed by Bragg area, as shown in Fig. 1(b) upper part, and a strong pulse normally incident on this structure from the left, as shown by the green arrow in Fig. 1(b). The front of the pulse will be reflected from the Bragg area, and as a result, an interference between the forward-propagating and backward-propagating parts of the pulse will occur in the above-Bragg area in immediate vicinity of above-Bragg/Bragg boundary. In this vicinity, high values of the squared electric field $E^2$ will lead to photoexcitation of carriers from valence to conduction zone and rise of the density of almost-free carriers (plasma), denoted by $\rho$ in Fig. 1(b). These carriers will provide negative contribution to the susceptibility of the material and reduce the refractive index $n_i$, transforming part of the above-Bragg area to Bragg area. Therefore the reflecting boundary of the above-Bragg and Bragg areas will dynamically shift to the left.

After that, the central part of the pulse will arrive to this new boundary and will be reflected from it, as depicted in the central panel of Fig. 1(b). By a similar process, photoinduced almost-free carriers will provide a negative contribution to the refractive index and transform another part of the formerly above-Bragg area to Bragg area, further shifting the boundary. A similar process will happen with the tail of the pulse, as shown in Fig. 1(b) lower part.

The net result of this process is the motion of the reflecting above-Bragg/Bragg boundary, which provides a moving "mirror" and can lead to pulse compression. The velocity of the motion can be close to $c$, since what moves is not matter but the position where a resonance condition $n_i=\lambda/(4d_i)$ is met. Note that the photoexcitation-induced "mirror" motion is governed by the same pulse which is reflected and compressed. 

\section{Theoretical and numerical model}

For the simulation of light propagation, the PIGLET software was used \cite{piglet}. The numerical model of this software is discussed in detail in \cite{piglet_our} and is presented here in a shortened form.

\begin{figure}
\includegraphics[width=1.0\textwidth]{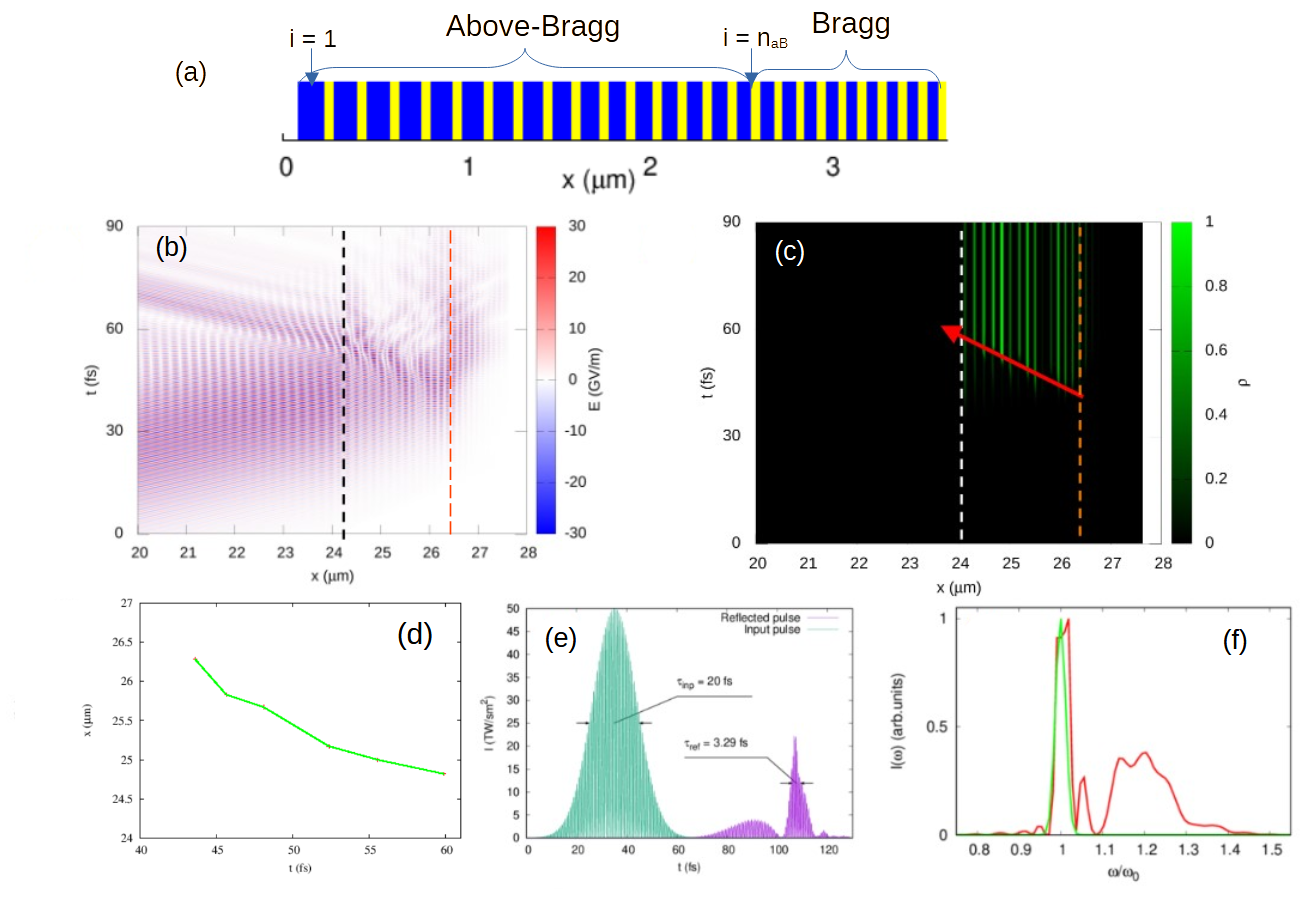}
\caption{Pulse compression in a silica/diamond structure. In (a), the considered structure of alternating fused silica (yellow) and diamond (blue) layers is presented, which are characterized by the parameters (for definition see text) $d_{aB,fs}=143.98$ nm, $d'_{aB,fs}=-3.77$ nm, $d_{B,fs}=68.56$ nm, and $d_{aB,d}=43.44$ nm, $d'_{aB,d}=-0.1$ nm, $d_{B,d}=41.37$ nm, respectively. The number of layers is $n_{aB}=21$ and $n_B=5$. In (b) and (c), the spatiotemporal maps of electric field and relative free-carrier density, respectively, are shown, with the front surface of the structure (vacuum/above-Bragg) and the initial position of the above-Bragg/Bragg boundary shown by white/black and orange dashed lines, respectively. In (d), the motion of the above-Bragg/Bragg boundary is depicted. In (e), the input and reflected pulse profiles are compared. In (f), the input (green) and reflected (red) spectra are shown. 50-TW/cm$^2$ input pulses with FWHM duration of 20 fs at 400 nm were considered.}
\end{figure}

We consider a normal incidence of linearly $y$-polarized pulse, which travel in the direction $+x$, on a plane-parallel layered structure of two dielectric materials. One-dimensional Maxwell equations for of electric $E_x(z,t)$ and magnetic $H_y(z,t)$ fields are written in the form
\begin{eqnarray} 
&&\epsilon_0\frac{\partial E_y(x,t)}{\partial t}=-\frac{\partial H_z(x,t)}{\partial x}-J_{ion}(x,t)-\frac{\partial P_y(x,t)}{\partial t}\\
&&\mu_0\frac{\partial H_z(x,t)}{\partial t}=\frac{\partial E_y(x,t)}{\partial x},
\end{eqnarray}
where $P_y(x,t)$ is the polarization, $J_{ion}(x,t)$ is the current which originates from the loss due to photoexcitation\cite{ion}: $J_{ion}(x,t)=N_0[1-\rho(x,t)]\Gamma(x,t)E_g/E_y(x,t)$, $N_0$ is the concentration of the entities (e.g. atoms or molecules) which can be ionized, $\rho(x,t)$ is the relative density of the ionized entities, $\Gamma(x,t)$ is the ionization rate, and $E_g$ is the bandgap. The  motion of the conduction-band electrons is characterized by the polarization
\begin{equation}
P_y(x,\omega)=E_y(x,\omega)\left[\epsilon_\infty-\frac{N_0\rho e^2}{\epsilon_0m_e\omega(\omega-i\nu)}\right]
\end{equation}
where $e$ is the electron charge, $\nu$ is the decay rate of the electron motion, and $m_e$ is the effective electron mass. Here, we neglected chromatic group velocity dispersion and Kerr polarization due to short propagation distance.

The ionization rate $\Gamma(x,t)$ in the ADK approximation is expressed \cite{meth_adk}(in atomic units) as 
  \begin{equation}
    \Gamma(x,t)=\left(\frac{3e'}\pi\right)^{3/2}\frac1{3n^3(2n-1)}\left(\frac{4e'}{(2n-1)|E(x,t)|}\right)^{2n-1.5}\exp\left(-\frac{2}{3n|E(x,t)|}\right),
  \end{equation}
where $e'=2.71828\dots$ and $n=1/\sqrt{2E_g}$. This rate is corrected by a prefactor which fits the damage threshold (defined as plasma frequency $\omega_p=\sqrt{N_0e^2\rho/(\epsilon_0m_e)}$ being equal to the pump frequency\cite{app_our}) to the experimental values.
The pulse propagation is described using the standard Yee-cell (leapfrog) one-dimensional FDTD approach\cite{FDTD}, with the boundary-condition-type input for the incident pulse, and absorbing boundary conditions.

For the numerical simulations, we have considered interchanging layers of fused silica and diamond. We used the following values of the above parameters for diamond: $E_g=5.5$ eV, $m_e=0.48m_0$ where
$m_0$ is the mass of a free electron, $N_0=1.75\times10^{29}$ m$^{-3}$, $\epsilon_0=5.85$, and $\nu=0.1$ fs$^{-1}$, and for fused silica:  $E_g=8.9$ eV, $m_e=0.43m_0$, $N_0=2.3\times10^{28}$ m$^{-3}$, $\epsilon_0=2.127$, and $\nu=0.1$ fs$^{-1}$.

\section{Results and discussion}

First we consider the structure depicted in Fig. 2(a). Initially its above-Bragg area consists of $n_{aB}$=21 layers of fused silica with thicknesses $d_i=d_{fs,aB}+id'_{fs,aB}$, alternating with the same number of diamond layers with thicknesses $d_i=d_{d,aB}+id'_{d,aB}$. The Bragg area consists of $n_{B}=4$ pairs of layers with thicknesses $d_{fs,B}$ for fused silica and $d_{d,B}$ for diamond. The structure was carefully optimized in terms of its parameters to provide the best compression; the parameters given in the caption indicate that in the above-Bragg area the thicknesses decrease with $i$, as shown in Fig. 2(a). We consider 50-TW/cm$^2$ input pulses with full-width half-maximum (FWHM) duration of 20 fs at 400 nm. 

\begin{figure}
\includegraphics[width=1.0\textwidth]{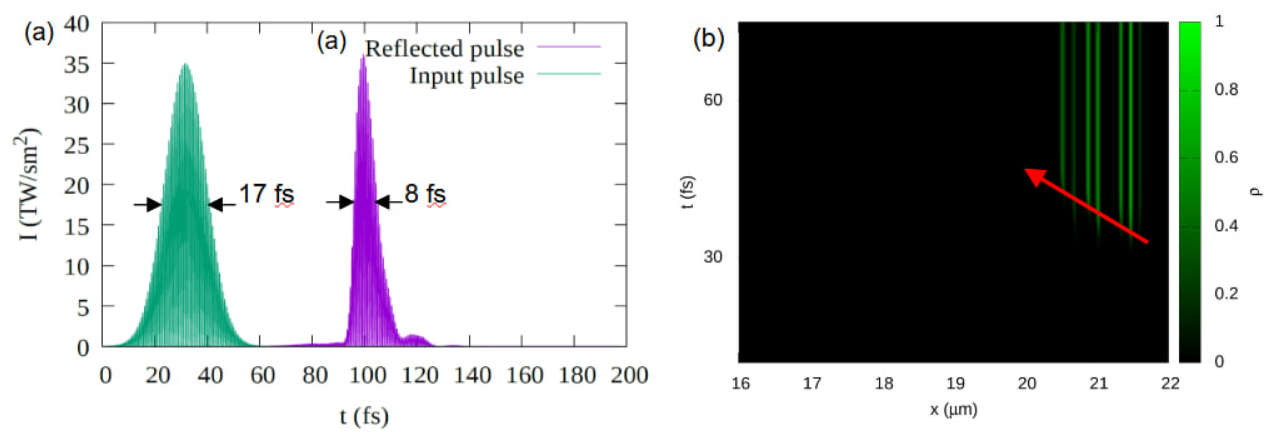}
\caption{Pulse compression in a silica/diamond structure with stronger reflected pulse. In (a), the input and reflected pulse profiles are compared, while in (b) spatiotemporal maps of relative free-carrier density is shown. The considered structure is characterized by the parameters (for definition see text) $d_{aB,fs}=111.24$ nm, $d'_{aB,fs}=-4.27$ nm, $d_{B,fs}=68.56$ nm, and $d_{aB,d}=67.06$ nm, $d'_{aB,d}=-2.56$ nm, $d_{B,d}=41.37$ nm, respectively. The number of layers is $n_{aB}=10$ and $n_B=5$. 35-TW/cm$^2$ input pulses with FWHM duration of 17 fs at 400 nm were considered.}
\end{figure}

In Fig. 2(b), the spatiotemporal map of the pulse propagation is presented. One can see that the pulse impinges on the surface of the structure (black dashed line) at around $t=35$ fs and is mostly transmitted into the above-Bragg area. It gets reflected at the boundary of the above-Bragg/Bragg areas (orange dashed line) and propagates back before emerging from the structure at around 57 fs with a significantly reduced duration. To understand the origin of compression, let us examine the spatiotemporal map of the induced relative plasma density $\rho$ shown in Fig. 2(c). One can see that the green vertical stripes, which indicate plasma, first occur at $t\simeq44$ fs near the above-Bragg/Bragg boundary. Note that diamond layers remain mostly not photoexcited due to lower photoionization rate. At later times, the plasma starts to occur for gradually smaller $x$ and reduces the refractive index. This corresponds to gradual conversion of above-Bragg area to Bragg area, as schematically presented in Fig. 1(b), and photoinduced motion of the reflecting above-Bragg/Bragg boundary in the $-x$ direction. The red arrow in Fig. 2(c) additionally indicates this motion. 

To get a quantitative information about the transformation of the above-Bragg area to Bragg area, we have recorded the time momenta at which the Bragg condition $n_i=\lambda/(4d_i)$ becomes fulfilled (with 10\% accuracy) in the fused-silica layers. The results are shown in Fig. 2(d), where one can see the monotonous motion of the above-Bragg/Bragg area in $-x$ direction. The resultant compression of the pulse is depicted in Fig. 2(e), with 6-fold pulse compression from 20 fs to 3.3 fs FWHM. In principle, during the reflection from a counter-propagating mirror, not only the pulse duration but also the optical period should reduce, corresponding to the spectral blue-shift. In Fig. 2(f), we test the presence of such shift by comparing the input and the reflected spectra, which shows a clear formation of a broad blue-shifted peak up to 1.4 times the incident frequency.

The well-pronounced and monotonous motion of the above-Bragg/Bragg boundary, pulse shortening, and the blue shift of the spectrum definitely indicate that the dynamical Bragg mirror is the main mechanism responsible for the pulse shortening in the considered case. We note parenthetically that using the velocity of the above-Bragg/Bragg boundary as derived from Fig. 2(d), one might hope to estimate the pulse compression and the frequency blue-shift, which in the moving-mirror limit should be given by the same factor $1/(1-v/c)$. Such estimate does not work well in the considered case of the photoionization-induced motion of the reflecting boundary, due to several effects, such as discreteness of the layered medium, intricate spatiotemporal field dynamics, and above all, changes in spectrum [Fig. 2(f)] which lead to modification of Bragg condition.  

The reflected pulse in the case considered in Fig. 2 has significantly lower peak intensity than the input pulse. However, for another parameters of the structure, as indicated in the caption of Fig. 3, we also predict a higher peak intensity of the reflected pulse. The increase of the intensity is minor, however, we emphasize that intensity increase is counteracted by the energy absorption for photoionization, therefore even such a minor intensity increase additionally indicates the action of the dynamical Bragg mirror mechanism. In this case, we predict only twofold compression from 17 fs to 8 fs FWHM. The spatiotemporal map of the relative plasma density, exhibited in Fig. 3(b), again shows gradual photoionization starting at the above-Bragg/Bragg boundary and moving in the $-x$ direction, in accordance with Fig. 1(b).

\begin{figure}
\includegraphics[width=1.0\textwidth]{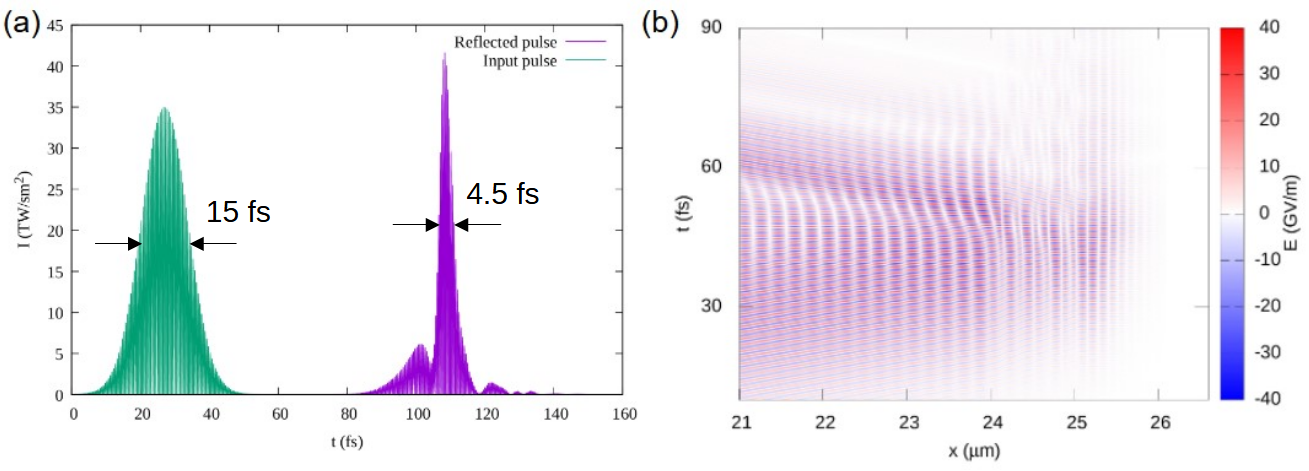}
\caption{Pulse compression at 300 nm. In (a), the input and reflected pulse profiles are compared, while in (b) spatiotemporal maps of electric is shown. The considered structure is characterized by the parameters (for definition see text) $d_{aB,fs}=94.27$ nm, $d'_{aB,fs}=-4.38$ nm, $d_{B,fs}=50.50$ nm, and $d_{aB,d}=55.11$ nm, $d'_{aB,d}=-2.56$ nm, $d_{B,d}=29.52$ nm, respectively. The number of layers is $n_{aB}=10$ and $n_B=5$. 35-TW/cm$^2$ input pulses with FWHM duration of 15 fs at 300 nm were considered.}
\end{figure}

Finally, we establish the feasibility of the proposed mechanism for even shorter wavelength of 300 nm in the near-uv region. In Fig. 4(a), we compare the input and reflected pulses, showing more than 3-fold compression from 15 fs to 4.5 fs, accompanied with notable increase of peak intensity. The spatiotemporal map of the electric field, shown in Fig. 4(b), shows the formation of the interference pattern near the initial above-Bragg/Bragg boundary.

For the experimental implementation of this proposal, several issues will need to be overcome. First, manufacturing of the multilayer structure needs to be performed with sufficient precision. Second, the proposed mechanism implies operation just below the damage threshold, which requires high stability of the input pulse intensity. Third, for high repetition rate heat accumulation will eventually lead to damage, which requires solutions such as laterally moving sample. However, we hope that these limitations will not pose insurmountable challenges.

\section{Conclusion}

We have designed a novel approach to pulse compression, based on the photoexcitation-governed counter-propagating motion of the reflecting above-Bragg/Bragg boundary. We predict pulse compression by factors up to 6, e.g. from 20 fs to 3.3 fs for intense 400-nm pulses. The feasibility of this mechanism for different input parameters and structures is demonstrated.

We note that the proposed design possesses several degrees of freedom which can be used to improve the performance, such as optimization of the layer thicknesses, choice of the layer materials, as well as control of the above-Bragg/Bragg boundary by one pulse while compressing another pulse. These possibilities will become the subject of consequent studies.


\begin{thebibliography}{5}

\bibitem{zewail}A. Zewail, "Femtochemistry: Atomic-scale dynamics of the chemical bond," J. Phys. Chem. A {\bf 104}, 5660–5694 (2000).

\bibitem{remacle}F. Remacle and R. D. Levine, "An electronic timescale in chemistry," PNAS {\bf 103}, 6793–6798 (2005).

\bibitem{dima} D. A. Zimin {\it et al.}, "Ultra-broadband all-optical sampling of optical waveforms", Sci. Adv. {\bf 8}, 1029 (2022). DOI:10.1126/sciadv.ade1029

\bibitem{a-streaking} R. Kienberger {\it et al.}, "Atomic transient recorder", Nature {\bf 427}, 817 (2004).
URL https://doi.org/10.1038/nature02277

\bibitem{nps} S. Sederberg {\it et al.}, "Attosecond optoelectronic field measurement in solids",
Nature Communications {\bf 11}, 430 (2020). URL https://doi.org/10.1038/s41467-019-14268-x

\bibitem{remacle2}F. Remacle, M. Nest, and R. D. Levine, "Laser Steered Ultrafast Quantum Dynamics of Electrons in LiH," Phys. Rev. Lett. {\bf 99}, 183902 (2007).

\bibitem{mod}T. Mocek, J. Polan, P. Homer, K. Jakubcza1, B. Rus, I. J. Kim, C. M. Kim, G. H. Lee, C. H. Nam, V. Hajkova, J. Chalupsky, and L. Juha, "Surface modification of organic polymer by dual action of extreme ultraviolet/visible-near infrared ultrashort pulses,"
Journal of Applied Physics {\bf 105}, 026105 (2009).


 \bibitem{Ell} R. Ell, U. Morgner, F. X. Kärtner, J. G. Fujimoto, E. P. Ippen,  V. Scheuer, G. Angelow, T. Tschudi, M. J. Lederer, A. Boiko, and B. Luther-Davies, "Generation of 5-fs pulses and octave-spanning spectra directly from a
Ti:sapphire laser," Opt. Lett. {\bf 26}, 373-375 (2001).

\bibitem{Brida} D. Brida, C. Manzoni, G. Cirmi, M. Marangoni, S. Bonora, P. Villoresi, S. De Silvestri, and G. Cerullo, "Few-optical-cycle pulses tunable from the visible to the mid-infrared by optical parametric amplifiers," J. Opt. {\bf 12}, 013001 (2010).

\bibitem{Nisoli} M. Nisoli, S. De Silvestri, O. Svelto, R. Szipocs, K. Ferencz, C. Spielmann, S. Sartania, and F Krausz, 
"Compression of high-energy laser pulses below 5 fs," Opt. Lett. {\bf 22}, 522-524 (1997).

\bibitem {last} M. Galli, V. Wanie, D. P. Lopes, E. P. Månsson, A. Trabattoni, L. Colaizzi, K. Saraswathula, A. Cartella, F. Frassetto, L. Poletto, F. L\'egar\'e, S. Stagira, M. Nisoli, R. M. V\'azquez, R. Osellame, and F. Calegari, "Generation of deep ultraviolet sub-2-fs pulses," Opt. Lett. {\bf 44}, 1308-1310 (2019).

\bibitem{Varillas} R. B. Varillas, A. Candeo, D. Viola, M. Garavelli, S. De Silvestri,
G. Cerullo, and C. Manzoni, "Microjoule-level, 
tunable sub-10 fs by
broadband sum-frequency generation," Opt. Lett. {\bf 39}, 3849-3851 (2014).

\bibitem{short_hh}A. Husakou and J. Herrmann, "Soliton-effect pulse compression in the single-cycle regime in broadband dielectric-coated metallic hollow waveguides," Opt. Express {\bf 17}, 17636-17644 (2009).

\bibitem {Brahms} C. Brahms, F. Belli, and J. C. Travers, "Infrared attosecond field transients and UV to IR few-femtosecond pulses generated by high-energy soliton self-compression," Phys. Rev. Res. {\bf 2}, 043037 (2020).

\bibitem{our_short} Anton Husakou, Ihar Babushkin, Olga Fedotova, Ryhor Rusetsky, Tatsiana Smirnova, Oleg Khasanov, Alexander Fedotov, Usman Sapaev, and Tzveta Apostolova, "Tunable in situ near-UV pulses by transient plasmonic resonance in nanocomposites," Opt. Express {\bf 31}, 37275-37283 (2023)

\bibitem{hassan} M. Hassan, T. Luu, A. Moulet, O. Raskazovskaya, P. Zhokhov, M. Garg, N. Karpowicz, A. M. Zheltikov, V. Pervak, F. Krausz, and E. Goulielmakis,  "Optical attosecond pulses and tracking the nonlinear response of bound electrons," Nature {\bf 530}, 66-70 (2016).

\bibitem{hoegen}A. von Hoegen, R. Mankowsky, M. Fechner,  M. F\"orst, and A. Cavalleri, "Probing the interatomic potential of solids with strong-field nonlinear phononics," Nature {\bf 555}, 79-82 (2018).

\bibitem{schultze} M. Schultze, E. M. Bothschafter, A. Sommer, S. Holzner, W. Schweinberger, M. Fiess, M. Hofstetter, R. Kienberger, V. Apalkov, V. S. Yakovlev, M. I. Stockman, and F. Krausz, "Controlling dielectrics with the electric field of light," Nature {\bf 493}, 75-78 (2013).

\bibitem{hanus} V. Hanus, V. Csajb\'ok, Z. P\'apa, J. Budai, Z. M\'arton, G. Kiss, P. S\'andor, P. Paul, A. Szeghalmi, Z. Wang, B. Bergues, M. Kling, G. Moln\'ar, J. Volk, and P. Dombi, "Light-field-driven current control in solids with pJ-level laser pulses at 80-MHz repetition rate," Optica  {\bf 8}, 570-576 (2021).


\bibitem{schiffrin} A. Schiffrin, T. Paasch-Colberg, N. Karpowicz, V. Apalkov, D. Gerster, S. M\"uhlbrandt, M. Korbman, J. Reichert, M. Schultze, S. Holzner, J. V. Barth, R. Kienberger, R. Ernstorfer, V. S. Yakovlev, M. I. Stockman, and F. Krausz, "Optical-field-induced current in dielectrics," Nature {\bf 493}, 70-74 (2013).

\bibitem{plasma}F. Quere, C. Thaury, P. Monot, S. Dobosz, Ph. Martin, J.-P. Geindre and P. Audebert, "Coherent Wake Emission of High-Order Harmonics from Overdense Plasmas", Phys. Rev. Lett. {\bf 96}, 125004 (2006).

\bibitem{piglet} https://github.com/AntonHusakou/PIGLET

\bibitem{piglet_our} A. Husakou, Z. Ruziev, K. Koraboev, F. Morales, M. Richter, and K. Yabana, "Benchmarking of analytical photoionization models for solids using photoionization-induced reflection",
Phys. Rev. A {\bf 110}, 063511 (2024).

\bibitem{ion} P. J\"urgens, B. Liewehr, B. Kruse, C. Peltz, D. Engel, A. Husakou, T. Witting, M. Ivanov, M. J. J. Vrakking, T. Fennel, and A. Mermillod-Blondin, "Origin of strong-field-induced low-order harmonic generation in amorphous quartz", Nature Physics {\bf 16}, 1035–1039 (2020). 

\bibitem{meth_adk}M. V. Ammosov, N. B. Delone, and V P. Krainov, “Tunnel ionization of complex atoms and atomic ions in an electromagnetic field,” Sov. Phys. JETP {\bf  64}, 1191–1196 (1986).

\bibitem{app_our} A. Husakou, I. Babushkin, O. Fedotova, R. Rusetsky, T. Smirnova, O. Khasanov, A. Fedotov, U. Sapaev, and T. Apostolova, "Tunable in situ near-UV pulses by transient plasmonic resonance in nanocomposites," Opt. Express {\bf 31}, 37275-37283 (2023).

\bibitem{FDTD} K. S. Yee, "Numerical solution of initial boundary value problems involving maxwell's equations in isotropic media", IEEE Transactions on Antennas and Propagation {\bf 14}, 302 (1966).


\end{thebibliography}
\end{document}